\documentclass[manuscript]{aastex}

\usepackage{lscape}

\usepackage[usenames,dvips]{color}
\usepackage{graphicx}

\shorttitle{Outer gap accelerator
 closed  by magnetic pair-creation process}
\shortauthors{Takata J. et al. }

\begin{document}


\title{Outer gap accelerator closed by magnetic pair-creation process}


\author{J. Takata\altaffilmark{1}, Y.Wang\altaffilmark{2} and
K.S. Cheng\altaffilmark{3}}
\affil{Department of Physics, University of Hong Kong,
Pokfulam Road, Hong Kong}

\altaffiltext{1}{takata@hku.hk}
\altaffiltext{2}{yuwang@hku.hk}
\altaffiltext{3}{hrspksc@hkucc.hku.hk}


\begin{abstract}
We discuss  outer gap closure mechanism in the trans-field direction 
with the magnetic pair-creation process near the stellar surface. 
The gap closure by the magnetic pair-creation is possible 
if some fraction of the pairs are produced with an outgoing  momentum. 
By assuming that  multiple magnetic  field will affect the local field near the
 stellar surface,  we show a specific magnetic field geometry 
near the stellar surface resulting in the outflow of the pairs. 
Together with the fact that the electric field is weak below null charge
 surface, the characteristic curvature photon energy emitted by incoming
 particles, which were accelerated in the outer gap, decreases
 drastically to $\sim 100$MeV near the
 stellar surface. We estimate
 the height measured from the last-open field line, above which 100~MeV
 photons is converted into pairs by the magnetic pair-creation.  
We also show the resultant multiplicity due to the magnetic
 pair-creation process  could acquire  $M_{e^{\pm}}\sim 10^4-10^5$. 
 In this model the
fractional outer gap size is proportional to $P^{-1/2}$.  The predicted
gamma-ray luminosity ($L_{\gamma}$) and the characteristic curvature photon
energy ($E_c$) emitted from the outer gap are proportional to $B^2P^{-5/2}$
and $B^{3/4}P^{-1}$ respectively.  This model also
predicts that  $L_{\gamma}$
and $E_c$ are related to the spin down power ($L_{sd}$)
or the  spin down age of pulsars ($\tau$) as
$L_{\gamma} \propto L_{sd}^{5/8}$ or  $L_{\gamma} \propto
\tau^{-5/4}$, and  $E_c \propto L_{sd}^{1/4}$ or $E_c \propto \tau^{-1/2}$
respectively.

\end{abstract}


\keywords{pulsars: general-- radiation mechanisms:non-thermal-- -gamma
rays:theory--magnetic field}



\section{Introduction}

The mechanism of  particle acceleration and high-energy emission
processes in the pulsar magnetospheres
are one of the unresolved physics of the pulsar activities. The
particle acceleration process and resultant high-energy
$\gamma$-ray  emission process have been discussed with the polar cap model
(Ruderman \& Sutherland 1975; Daugherty \& Harding 1982),
the slot gap model (Arons 1981; Harding, Usov \& Muslimov 2005; Harding
et al. 2008) and
the outer gap  model (Cheng, Ho \& Ruderman 1986a,b; Hirotani 2008;
Takata \& Chang 2009). The polar cap model
assumes the emission site  close to the stellar surface above the polar cap,
and the slot gap and the outer gap models assume the emission site in the
outer magnetosphere. The different
acceleration  models have predicted the different properties of the
$\gamma$-ray emissions from the pulsar magnetospheres.

The observations of the pulsar  emitting electromagnetic radiation
in  the high-energy
$\gamma$-ray bands have been facilitated by recent space and ground based
 telescopes.    In  particular,  the $Fermi$ $\gamma$-ray telescope
 has measured the  $\gamma$-ray emissions from  $\sim 46$  pulsars
 (Abdo et al. 2010, 2009a,b), including 21 radio-loud,  17 radio-quiet
and 8 millisecond  pulsars. In addition, $AGILE$
(Astro-rivelatore Gamma a Immagini LEggero) has
also reported the detection of $\gamma$-ray emissions from 4 new pulsars
with 4 candidates (Pellizzoni et al. 2009). In more higher energy regime,
 $MAGIC$ (Major Atmospheric Gamma Imaging
Cherenkov) telescope
has detected for the first time pulsed gamma-ray radiation above 25~GeV
from the Crab pulsar  (Aliu et al. 2008).
These observations will be useful to discriminate between
the emission models.  For example, the $Fermi$ telescope  has measured
 the  spectral properties  above 10~GeV  with a  better
sensitivity than $EGRET$.
 It was   found that the spectral shape of $\gamma$-ray emissions from
 the Vela pulsar is well fitted with a power low
(photon index $\Gamma\sim$1.5) plus exponential
 cut-off ($E_{cut}\sim 3$~GeV) model. The discovered exponential
 cut-off feature predicts that  the emissions
from the outer magnetosphere  (Abdo et al. 2009c) is more favored   than the
 polar cap region (Daugherty \& Harding 1996), which predicts a super
exponential cut-off with the magnetic pair-creation.
Furthermore, the detection of the radiation above 25~GeV
bands associated with the Crab pulsar has also predicted the high-energy
emission in  the outer magnetosphere (Aliu et al. 2008).

The pulse profiles observed by the $Fermi$ telescope allow us to study the
site of the $\gamma$-ray emissions in the pulsar magnetosphere. Venter
et al. (2009) fitted the pulse profiles of the 8 millisecond pulsars with
the geometries predicted by the different emission models.
They showed that most of the pulse profiles  can be best fit with the
outer gap   (Takata et al. 2007; Takata \& Chang 2007;
Tang et al. 2008)  or the two pole caustic
(Dyks \& Rudak 2003; Dyks et al.  2004) geometries, which have
a slab like  geometry along the last-open field lines.
  However, they also
found that the pulse profiles  of  two out of
eight   millisecond pulsars cannot be fitted by either
 the  geometries with the outer gap or  the caustic models.  They proposed
 a pair-starved polar cap model, in which the multiplicity of the pairs
is not high enough to completely screen the electric field above the
polar cap,  and the particles are continuously accelerated up to high
altitude over full open field line region.

The increase of the $\gamma$-ray pulsars
 allows us to perform a detail  statistical
study of the $\gamma$-ray pulsars. In particular,  the $Fermi$
$\gamma$-ray pulsars including millisecond pulsars
will reveal the relation between the $\gamma$-ray
luminosity ($L_{\gamma}$) and the spin down power ($L_{sd}$),
 for which $L_{\gamma}\propto L_{sd}^{\beta}$ with
 $\beta\sim 0.5-0.6$ was predicted by $EGRET$ measurements (Thompson 2004).
Also, the $Fermi$ $\gamma$-ray pulsars will enable us to discuss the
general trend of the relation among
 the spectral properties of the $\gamma$-ray emissions
(e.g. the cut-off energy and photon index)  and the pulsar parameters
(e.g. rotation period and surface magnetic field).
Together with the observed pulse profiles and the spectra,
these  general properties of
the $\gamma$-ray emissions
 will discriminate between the $\gamma$-ray emission models
in the pulsar magnetospheres.

In this paper, we discuss the $\gamma$-ray emissions from the
outer gap accelerator. We propose a new outer gap closure mechanism 
 by the magnetic pair-creation process near the stellar
surface. The pairs produced by the magnetic pair creation will be able
to close the gap  if the sufficiently strong surface multiple  
field exists and  affects  the dipole field near the surface. 
In section~\ref{closure},  we first summarize results of 
gap closure process  by   photon-photon pair-creation process, 
 and  then we discuss our  new gap closure mechanism by  the magnetic
pair-creation process. In section~\ref{outergap}, we describe 
 the model predictions of the
 properties of the $\gamma$-ray emissions.  In section~\ref{discussion}, 
we will compare
the model predictions with the results of the  $Fermi$ observations. We
also discuss applicability of  our model. In
section~\ref{conclusion},  we will
 summarize our gap  closure model and predictions for the outer gap
 accelerator.

\section{Gap closure mechanism}

\label{closure}

\subsection{Photon-photon pair-creation process}

\label{photon}
The outer gap accelerator model was proposed by Cheng, Ho and Ruderman 
(1986a,b), who argued that a large global current flow through the outer
magnetosphere causes a charge depletion from the Goldreich-Julian charge
density, which is defined by  $\rho_{GJ}=-\vec{\Omega}\cdot \vec{B}/2\pi
c$ (Goldreich \& Julian 1969) with $\vec{\Omega}$ being the vector of the
rotation axis and $\vec{B}$ the magnetic field. In the charge depletion
region, the non-corotational electric field along the magnetic field
 accelerates the charged particles, which result in the high energy 
 $\gamma$-ray emissions.  This non-corotational electric field could  
be screened out by the discharge of the copious electron and positron
pairs produced  by the pair-creation process of the $\gamma$-ray photons.  
The outer gap would be  completely 
screened out in the trans-field direction, 
where  the non-corotational  electric
 field perpendicular to the magnetic field in the
 poloidal plane is equal to zero.  This condition
implies that the total potential (corotational + non-corotational) 
field is continuously connected to the corotational field outside
 the  outer gap.  

Zhang \& Cheng (1997) discussed the gap closure mechanism  by
  the photon-photon pair-creation process between  the
high-energy $\gamma$-rays emitted in the gap
and the X-rays  coming  from the stellar surface.
They estimated the typical gap thickness from  the pair-creation condition 
$E_{\gamma}E_X\sim 2(m_ec^2)^2$, where $E_{\gamma}$ is the energy of the
emitted $\gamma$-ray photons in the outer gap and $E_{X}$ is the energy
of the soft-photons from the stellar surface, and they obtained 
 the fractional gap thickness  as
$f_{p}\equiv h_{\perp} (R_{lc}/2)/R_{lc}\sim
 5.5P^{26/21}B_{d,12}^{-4/7}$, where $R_{lc}$ is the light cylinder radius,
$h_{\perp}(R_{lc}/2)$ is the gap thickness in poloidal plane at
$r=R_{lc}/2$ and $B_{d, 12}$ is the global
stellar magnetic field in units of $10^{12}$~Gauss.

The outer gap closure mechanism with the photon-photon pair-creation
process in the trans-field direction 
have also been discussed by solving the electrodynamics in the
outer gap with 2-dimensional  and 3-dimensional geometry (Takata,
Shibata and Hirotani 2004; Hirotani 2006a,b; Hirotani 2008). 
For example, Hirotani
(2006a) demonstrated that the outer gap for the young pulsar, 
 the Crab pulsar,  is almost screened out  in the
trans-field direction at the fractional gap thickness of  $f\sim 0.2$
(figure 6 in Hirotani 2006a). For mature pulsars such like Gemiga, 
on the other hand,  Takata and Chang (2009) argued 
that the photon-photon pair-creation process will be insufficient in the
outer magnetosphere and
  the outer gap could  occupy entire region between the 
last-open field lines and the critical field lines that have the null
charge point at the light cylinder.

\subsection{New gap closure mechanism; magnetic pair-creation process 
near the stellar surface}
\label{magnetic}

In this paper, we propose  a possible  gap closure  mechanism, in which 
the magnetic par-creation process near the stellar surface supplies the 
electron and positron pairs to close the outer gap in the trans-field
direction.  
It has been proposed that the magnetic field near the stellar surface is
enhanced  by the strong multiple magnetic field, although the global
magnetic field is well described by the dipole field (Blandford et
al. 1983; Romani 1990; Ruderman 1991). The neutron star magnetic
 field will be  produced by a current flowing  the crust, which
 has a thickness of $\delta r\sim 1-3 \times 10^5~\mathrm{cm}\ll R_s$.
It was  suggested that the magnetic structure near the stellar
 surface is super position of  clumps resolved into multiples of
 characteristic  of the order of  $ R_s/\delta r$ (Arons 1993; Zhang \&
 Cheng 2003).
 If the clumps cover whole stellar surface, the strength of the
 stellar magnetic field is order of $B_s\sim (R_s/\delta r)^nB_d$, where
 $B_d$ is the strength of dipole magnetic field determined by the  observed
rotation period $P$ and the period derivative $\dot{P}$,
and $n=1$ and 2  represent
coherent and incoherent superposition of the magnetic momentum of the
clump, respectively. This model  indicates
the strength of the stellar magnetic field can take
easily $B_s\sim 10-100B_d$.
Therefore, even in the millisecond pulsar, the magnetic field near the
star surface will be close to  $B_s\sim 10^{11}$~Gauss, and the magnetic
pair-creation process will take place close to the stellar surface.

We will argue that the $\gamma$-rays emitted near the stellar surface 
is converted into the pairs via the magnetic pair-creation process above 
the height ($h_{\perp,m}$) measured from the last-open field line. In
equation~(\ref{height}),  we will estimate the height,  $h_{\perp,m}$,
above which the pair-creation process  takes  place. 
The important conditions for closing the outer gap 
by the pairs produced by the magnetic pair-creation process near the
stellar surface are as follows; 
 (1) the gap was  not closed by the
photon-photon pair-creation process  
below the height $h_{\perp,m}$ (c.f. section~\ref{limit}),  
 (2) the local magnetic field lines near the stellar surface
 is bending away from the last-open field 
lines (see below and  Figure~\ref{Pulsar}) due to the strong multiple
fields, and (3) some of magnetic pairs migrate into outer magnetosphere. 
The second condition is required to produce outflows of the magnetic
pairs. 

\begin{figure}
\epsscale{.50}
\plotone{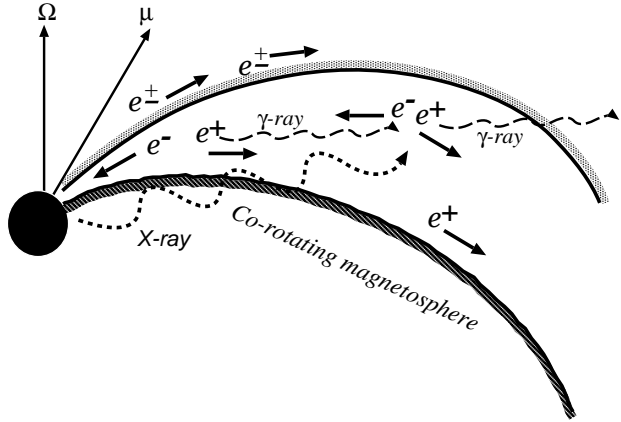}
\caption{Schematic view of the outer gap accelerator.  In the outer
 magnetosphere, the photon-photon pair-creation process produce the
 pairs in the gap uniformly. The pairs created by the magnetic pair-creation
 process close the outer gap in the outer magnetosphere. 
The favorable magnetic structure near the stellar surface
 is depicted in Figure~\ref{Pulsar}.}
\label{Pulsar1}
\end{figure}

Figure~\ref{Pulsar1} and~\ref{Pulsar} represent the schematic view 
of the outer magnetosphere and of the magnetic structure near the
stellar surface, which is favored  in this study, respectively. 
 In the outer magnetosphere, the photon-photon pair-production process
create  pairs, which are  separated by the accelerating
electric field. Inward propagating  particles will 
emit the $\gamma$-ray photons toward the strong magnetic
field region near the stellar surface. 
As demonstrated by the electrodynamic study (e.g. Hirotani 2006a), the
electric field below  null charge
surface are significantly reduced  by the pairs and arises
with a very weak field. Below the null charge surface,  therefore,
the curvature energy loss will not be able to be compensated
by the acceleration of
the electric field in the gap (c.f. section~\ref{valid}), and the incoming 
particles loose their energy by the curvature radiation. It 
is interesting to note that there is a minimum energy of the curvature 
photons, which does not depend on any pulsar parameters
 and the curvature radius of the local 
magnetic field. Assuming that the curvature loss dominates the energy
gain due to the acceleration by the electric field,
the evolution of the Lorentz factor may be  described as
\begin{equation}
m_ec^2\frac{d\gamma}{dt}=-\frac{2}{3}\gamma^4\frac{e^2c}{s^2},
\label{loss}
\end{equation}
where $s$ is the curvature radius.  Close to the null charge surface, 
the incoming  particles lose their energy 
with a time scale smaller than the time
scale of travelling to the stellar surface, 
 because the Lorentz factor is high enough. As decreasing the
Lorentz factor of the particles, the curvature energy loss time scale
becomes comparable to the crossing time scale of $dt\sim s/c $. 
In such a case,  equation of
motion~(\ref{loss}) implies  $1/\gamma^2-1/\gamma_0^2\sim 4e^2/3m_ec^2$
where  $\gamma_0$ is the initial Lorentz
factor. If $\gamma<<\gamma_0$,  the typical Lorentz factor of the
particles  below the null charge surface becomes 
\begin{equation}
\gamma\sim \left(\frac{3m_ec^2s}{4e^2}\right)^{1/3}\sim 3\times 10^6
s_7^{1/3},
\label{lorent}
\end{equation}
where $s_7$ is the curvature radius 
in units of $10^7$~cm. 

 The Lorentz factor (\ref{lorent}) gives the  minimum energy 
of the curvature photons, which does not
depend on any pulsar parameters, that
\begin{equation}
E_{min}\sim \frac{3}{4}\frac{\hbar\gamma^3c}{s}\sim
 \frac{9m_ec^2}{8\alpha_f}
\sim 77~\mathrm{MeV},
\end{equation}
where $\alpha_f$ is the fine structure constant. We expect  that the
incoming electrons will mainly emit the curvature photons with an energy
$E_{min}\sim 100$~MeV between the stellar surface and the null charge
surface.  Some of 100~MeV photons emitted below the null charge surface 
pass through vicinity of the stellar surface.  
  Applying the static dipole field geometry  and $P=0.1$,  for example, 
 the trajectory of the 
100~MeV photons emitted below  $r\sim 10^7$~cm on the last-open
filed  line will cross the magnetic pole below 
$r\sim 2\times 10^6$~cm. Therefore,  
it is expected almost all 100~MeV photons emitted below  $r\sim 10^7$~cm 
will be converted into pairs with a strong magnetic field of 
$B\sim10^{11}- 10^{13}$~Gauss.  
We find that  an incoming  particle emits
about $N\sim 2\times 10^4 s_7^{-2/3}$ photons  before reaching the stellar 
surface below  $r\sim 10^7$~cm . 

For the millisecond pulsars, the 100~MeV curvature photons could be
converted into the pairs with  a strong multiple  magnetic field, which could 
acquire $B_s\sim 10^{11}$~Gauss,  at very close to
the stellar surface. With the static dipole field geometry, 
the curvature photons emitted blow $r\sim3\times 10^6$~cm is propagating
toward the magnetic pole  for the typical rotation period 
of $P=4$~ms. An incoming particle will emit   
$N\sim  10^4 s_6^{-2/3}$ of $\sim 100$~MeV photons below $r\sim 3\times
10^6$~cm 

For the canonical pulsar, the pairs created 
by the magnetic pair-creation process of 100~MeV photons can also emit the soft
$\gamma$-rays via the synchrotron radiation, which may further generate 
 new pairs.   With the typical Lorentz factor of $\Gamma\sim 100$, 
one pair emits about 1$\sim$10 synchrotron photons 
with a typical energy of  $E_{syn}\sim 5(\Gamma/100)^2(\sin\theta_a/0.05)
(B/5\cdot 10^{11}~\mathrm{G})$~MeV.  This $\sim$5~MeV synchrotron photons
also could be converted into pairs with a strong
local magnetic field $10^{12}\sim 10^{13}$~Gauss near the stellar
surface, where the magnetic field will be enhanced by the multiple
magnetic field.  This implies  
 the resultant multiplicity of  an incoming particles, 
which was accelerated in the outer
gap, could be  $M_{e^{\pm}}\sim10^5 s^{-2/3}_7$ for the canonical pulsars.

The mean free path of the magnetic pair-creation process
of the photon with the energy $E_{\gamma}$ is described
as  (Erber 1966;Ruderman \& Surtherland 1975)
\begin{equation}
l_m=\frac{4.4}{\alpha_f}\frac{\hbar}{m_ec}\frac{B_q}{B_{\perp}}
\exp\left(\frac{4}{3\chi}\right),
\end{equation}
where $\chi=E_{\gamma}B_{\perp}/2m_ec^2 B_q$, $B_q=4.4\times 10^{13}$~Gauss and
$B_{\perp}=B\sin\theta_a$ with $\theta_a$ is the angle between
the direction of the propagating for the photon and the magnetic field.
The exponential dependency of the mean-free path on the photon energy
$E_{\gamma}$ and the angle $\theta_a$ implies that most of emitted
photons will be converted into pairs if the  condition
that $E_{\gamma}B_{\perp}/(2 m_ec^2B_q)\sim \chi $ is satisfied.
Using   the magnetic pair-creation
 condition,  we can estimate height measured from the last open field
line, above which  the magnetic
pair-creation process of the photons with $E_{min}\sim 100$~MeV
 becomes important process. 
The collision angle $\theta_a$  
is approximately
described as
\begin{equation}
\sin\theta_a\sim\frac{\ell}{s}\sim\sqrt{\frac{2h_{\perp}}{s}},
\end{equation}
where $\ell\sim \sqrt{2h_{\perp}s}$ is the propagating distance of a
photon from the emission point, and $h_{\perp}$ is height measured from
the last-open field line. The magnetic pair-creation condition
implies that  the curvature photons with an energy 
$\sim 100$~MeV  will be converted into the pairs above the height  
\begin{equation}
h_{\perp,m}(R_i)\sim 10^4\chi_{-1}^{2}B_{m,12}^{-2}s_7
~\mathrm{cm},
\label{height}
\end{equation}
where $R_i $ expresses  the critical radial distance below 
which the magnetic pair-creation process becomes to be important
 and it will be $R_i\sim 2-3R_s$ for the canonical pulsars and $R_i\sim R_s$ 
for the millisecond pulsars. In addition, 
$\chi_{-1}=\chi/0.1$, $B_{m,12}$ is the strength of the  magnetic
field at the pair-creation position in units of $10^{12}$~Gauss, and
we used $E_{min}=100~\mathrm{MeV}$.  Rescaling the thickness at the
stellar radius using the magnetic flux conservation,
we obtain $h_{\perp,m}(R_s)\sim (R_s/R_i)^{3/2} h_{\perp,m}(R_i)$ with $R_s$
being the stellar radius.

As we have discussed above, the incoming particles will  create pairs above
the height $h_{\perp,m}$, which is  described by
equation~(\ref{height}), via the magnetic pair-creation process 
near the stellar surface, and  the multiplicity could acquire 
  $M_{e^{\pm}}\sim 10^4-10^5$ for the
canonical pulsars and $\sim 10^4$ for the millisecond pulsars.  
Although it is expected most of the created pairs have inward momentum 
and migrate toward the star, it may be possible 
that small  fraction of the created pairs are produced 
with an outgoing  momentum and migrate into  outer magnetosphere, as
discussed below. In fact, about $\sim 10$ pairs  out of 
$M_{e^{\pm}}\sim 10^4-10^5$ will be enough to close the outer gap  
 accelerator in the trans-field direction everywhere in the outer 
magnetosphere.

The magnetic pairs would  be produced with an outgoing  momentum if the 
magnetic structure near the stellar surface is affected by the strong 
 multiple field,  as illustrated in Figure~\ref{Pulsar}, which
shows how some pairs can acquire the outgoing momentum due to the
geometry of the local magnetic field lines.  In fact, it is required
that the magnetic field (thick dashed-lines) 
on the pair-creation points are bending away
from the last-open field line due 
 to  a strong multiple field (solid-dashed line).  The solid
curved lines represent the global field which are not affected by the
local magnetic field, and  thin dashed-lines represent the  geometry
without the multiple  field.  The 
100~MeV curvature photons (solid arrows) emitted by the incoming
particles, which were accelerated inside the outer gap, are converted into the 
pairs via the magnetic pair-creation above 
height,  $h_{\perp,m}$ described by equation~(\ref{height}).
 For the canonical pulsars, the synchrotron
radiation of the created pairs will produces 
  $\sim$5~MeV photons (dashed arrows), which 
could be converted into  new pairs by 
 a strong magnetic field near the stellar surface.  If the direction of
 the magnetic field at the pair-creation positions is bending away from
 the last-open field line, it
 is possible that 
the collision angle (denoted as $\alpha$ in the Figure~\ref{Pulsar}) 
between the photons and the magnetic field line becomes larger than $\alpha\ge
90^{\circ}$, implying the created pair has the outgoing  momentum.   

We note that (1) because the position of the pair-creation point 
is determined by the value $B\sin\alpha$, and (2) because the collision 
angle changes from a smaller value to a larger value 
as the photon propagate toward the stellar surface, the position with
the collision angle smaller than $90^{\circ}$ is preferentially chosen
as the pair-creation point,  although $\sin \alpha$
gives the same values with, for example,  $\alpha=80^{\circ}$ and  
$110^{\circ}$. However, the magnetic field also increases as the photons
propagate toward the stellar surface, implying 
the strength of the magnetic field is bigger for the position with 
$\alpha=110^{\circ}$ than that with $\alpha=80^{\circ}$. As a result, it
will be possible that some photons do not have  enough energy to produce the
pairs with the magnetic field at the position with $\alpha=80^{\circ}$, 
but do to produce the pairs  at the position with
$\alpha=110^{\circ}$. On these ground,  some created pairs will be able
to have the outgoing  momentum, if the local magnetic field
is bending away from the last-open field line. 

With the magnetic field structure bending away from the last-open field
line, the incoming particles from the outer gap could  emit the 
100~MeV curvature photons  toward lower altitude, and therefore could 
make  pairs at lower altitude, implying the magnetic pair-creation 
process could  occur inside the gap if the inner boundary 
is located close to the stellar surface. The created pairs will be
discharged by the small electric field along the magnetic field and this
discharge will partially screen the electric field. Because the created
pairs lose their perpendicular momentum within very short distance, the
Lorentz factor after loosing the perpendicular momentum is $\Gamma\sim
1/\sin\theta_0\sim 10(\sin\theta_a/0.1)^{-1}$. The magnetic pairs will
screen the electric field near the inner boundary so that the potential
drop between the inner boundary and  the point, below which the magnetic
pair-creation process is occurred, is reduced to 
 $V\sim 5\times 10^6$~Volts.

If all field lines in the polar cap region 
are bending toward the last-open field line, 
the outgoing magnetic pairs are  not expected  because the collision angle 
is always $\alpha<90^{\circ}$. 
In such a case, all  pairs created by the magnetic pair-creation
process will have an inward momentum, implying  the photon-photon
pair-creation process will be only possible mechanism to close the 
outer gap.

\section{Outer gap closed by magnetic pair-creation process }
\label{outergap}
 Our gap closure process is summarized  as follows;
\begin{enumerate}
\item the incoming particles emit $\sim$100~MeV photons below the null
      charge surface,
\item  the $\sim$100MeV photons emitted toward the stellar surface will
      produce the pairs via the magnetic pair-creation process 
($M_{e^{\pm}}\sim 10^4-10^5$) above the height $h_{\perp,m}$ estimated 
      by equation (\ref{height}), and 
\item  if some fraction of the created pairs  will be produced with an
      outgoing momentum due to the geometry of the local magnetic field,
 then the outgoing  pairs close the outer gap 
in the trans-field direction everywhere in the outer magnetosphere.
\end{enumerate}
We will discuss  the applicability of the present model
in section~\ref{limit}.

Applying equation~(\ref{height}) as the typical thickness of the gap
rescaled at the radial distance $r=R_i$  with the dipole field geometry, 
we define  the fractional gap thickness on the stellar surface as
\begin{equation}
f_m\equiv \frac{h_{\perp,m}(R_s)}{r_p}\sim 0.25
K(\chi,B_m, s)P_{-1}^{1/2},
\label{frac}
\end{equation}
with
\begin{equation}
 K=\chi_{-1}^2B_{m,12}^{-2}s_7 \left(\frac{R_s}{R_i}\right)^{3/2}
\end{equation}

Applying the fractional gap thickness to the acceleration and curvature
 emission process beyond the null charge surface, we discuss the expected
 properties of the $\gamma$-ray radiation from the outer gap closed 
 by the magnetic pair-creation process.
The typical magnitude of the electric field in the gap beyond the
null charge surface  is given by
\begin{equation}
E_{||}(K,B_d,P)\sim \frac{f_m^2V_a}{R_{lc}}\sim 8.3\times 10^4 K^2
 B_{d,12}P^{-2}_{-1}~\mathrm{Volt/cm}.
\label{accfield}
\end{equation}
which can accelerate the electrons and positrons up to
\begin{equation}
\gamma(K,B_d)\sim \left(\frac{3s^2}{2e}E_{||}\right)^{1/4}
\sim 2.1\times 10^7K^{1/2}B_{d,12}^{1/4},
\end{equation}
where we used $s=R_{lc}$ in the outer magnetosphere. We find that
the maximum Lorentz factor does not depend on the rotational period.
The typical energy of the curvature radiation by the
accelerated particles in the outer magnetosphere is
\begin{equation}
E_{c}(K,B_d,P)=\frac{3}{4\pi}\frac{hc\gamma^3}{s}
\sim 0.55K^{3/2}B_{d,12}^{3/4}P^{-1}_{-1}~\mathrm{GeV}.
\label{cut-off}
\end{equation}
Because the efficiency of the emission  is significantly reduced above
 the energy $E_{c}$, we expect that the $\gamma$-ray
spectrum has the cut-off energy of $E_{c}$.

The total current flowing in the gap  is order of
$I_{gap}\sim f_m I_{GJ}$, where
$I_{GJ}=\pi B_dR_s/PR_{lc}$ is the Goldreich-Julain current.  Using
the total  potential drop in the gap of  $V_{gap}\sim f_m^2V_a$, we can
estimate the $\gamma$-ray luminosity as
\begin{equation}
L_{\gamma}(K,B_d,P)\sim I_{gap}V_{gap}
\sim 2\times 10^{33}K^3B_{d,12}^2P^{-5/2}_{-1}~\mathrm{erg/s}.
\label{lumi}
\end{equation}

\section{Application and discussion}
\label{discussion}
\subsection{Comparison with $Fermi$ observations}
\label{compari}
The present model predicts the cut-off energy and the luminosity of the
$\gamma$-ray emissions are related with the stellar magnetic field and
the rotation period as $E_c\propto B_d^{3/4}P^{-1}$ of equation~(\ref{cut-off})
and $L_{\gamma}\propto B_d^2P^{-5/2}$ of equation~(\ref{lumi}), respectively.
We  plots the our predictions with the
observations on the cut-off energy in Figure~\ref{cutoff}
 and on the luminosity in Figure~\ref{lumin}.  In the figures, the
circle and triangle symbols represent the radio-selected and
$\gamma$-ray selected $\gamma$-ray pulsars, and  the squares are the
millisecond pulsars.  We can see that the slope of the model  prediction is
consistent with the observations for both
canonical pulsars and  millisecond pulsars.

In the present outer gap model,
the properties of the $\gamma$-ray emissions depend on the local
parameters  $K(\chi, B_m, s)=\chi_{-1}^2B_{m,12}^{-2}s_7(R_s/R_i)^{3/2}$, which
is determined by the local magnetic structure.
In fact, we adopted $K=2$ for the canonical $\gamma$-ray pulsars
 and $K=15$ for
the millisecond pulsars in Figures~\ref{cutoff} and~\ref{lumin}.
The strength of the multiple field at a distance $\delta R$
from the stellar surface may be  expressed as $B_m\sim B_s[(\delta
r+\delta R)/\delta r]^{-(m+1)}$, where $B_s$ is the strength of the
multiple field at the stellar surface and $\delta r\sim 1-3 \times
10^5$~cm is the
 thickness of the crust, the index  $m$ is the multiplicity. If we consider the
localized dipole field ($m=2$), $B_m\sim B_s[(\delta
r+\delta R)/\delta r]^{-3}$, with $B_s\sim 10^{13}$~Gauss,
this localized dipole field becomes same order of magnitude
 with the global dipole field at about one stellar radius from
the stellar surface,
 because $B_m(r-R_s=R_s)\sim  B_s (\delta r/R_s)^3\sim 10^{-2}B_s$ with
 $\delta r\sim 2-3\times 10^5$~cm
and because the global dipole field becomes $B_d/2^3$ at
 $r-R_s=R_s$.  If the radial distance below which 
 the magnetic pair-creation process takes place  is $R_i=2R_s$, 
the strength of the  magnetic field at $R_i$ is $B_m\sim  10^{11}$~Gauss
 and the curvature  radius of the multiple
field will be $s\sim 10^6$~cm. This implies  the local
 parameter of $K=2^{-3/2}\chi^2_{-1}B_{m,12}^{-2}s_7\sim 3.5$, which explains
 $K\sim 2$ in Figures~\ref{cutoff} and ~\ref{lumin}.
For the millisecond pulsars, it is expected
 that the magnetic par-creation of the photons with the  energy
$E_{min}\sim 100$~MeV  is possible only at very close to the stellar
 surface ($r\sim R_s$), because the field strength is about
 three or four  order of magnitude smaller than that of the canonical
 pulsars. If we adopt
$B_s\sim 3\times 10^{10}$~Gauss as the strength of the  multiple field on the
stellar surface  and $s\sim 10^5$~cm as the curvature radius of the
multiple field, we obtain $K\sim 10$.

As Figures~\ref{cutoff} and~\ref{lumin} show, the present model predicts
that the local parameter $K(\chi, B_m, s)$ discriminates
the canonical and the millisecond
pulsars  as  the different populations
on the plots of  the cut-off energy $E_c$ versus   $B_d^{3/4}P^{-1}$
and of the luminosity  $L_{\gamma}$ versus  $B_d^2P^{-5/2}$.
 From equations~(\ref{cut-off}) and~(\ref{lumi}), on the other hand,
one can see that  $L_{\gamma}/E_c^2$ does not include the local factor
$K(\chi,B_m, s)$.  More strictly speaking,
the quantities  $L_{\gamma}/E_{c}^2$  carries away
the dependency on the gap thickness $f$, which depends on  the local
parameters $K$,  because $L_{\gamma}\propto f^3L_{sd}$ and
$E_c\propto f^{3/2}V_a^{3/4}s^{-1/4}$. We then  obtain the relation  that
\begin{equation}
\frac{L_{\gamma}}{E_{c}^2}\sim 6.6\times 10^{34}B_{d,12}^{1/2}P_{-1}^{-1}
~\mathrm{erg/s~GeV^2}.
\label{basiceq}
\end{equation}
 Figure~\ref{basic} compares the model prediction and
the observations  for each pulsar.
Figure~\ref{basic} shows that both canonical pulsars and millisecond
pulsars are consistent with  the line of $L_{\gamma}/E_c^2\propto
B_{d,12}^{1/2}P_{-1}^{-1}$.
This predicts that
 although there is a gap between the two
populations,  where no
$\gamma$-ray pulsars are plotted in Figure~\ref{basic},
 in fact two populations will be
continuously connected.  It is expected that  more $\gamma$-ray
pulsars having smaller  $B_{d,12}^{1/2}P_{-1}^{-1}$ than the present
canonical $\gamma$-ray pulsars will be discovered by $Fermi$
$\gamma$-ray
telescope. Those pulsars
 will be expected to distribute around solid
line in  Figure~\ref{basic},  and  two
 populations, i.e. the canonical and millisecond pulsars, will be
 continuously connected in the plot.

 Recasting the cut-off energy $E_c$  given by equation~(\ref{cut-off}) and
the $\gamma$-ray luminosity $L_{\gamma}$ given by equation~(\ref{lumi}) 
as a function
of the spin down age $\tau=P/2\dot{P}$, we  obtain
\begin{equation}
E_{c}\sim 7K^{3/2}B_{d,12}^{-1/4}\tau_3^{-1/2}~\mathrm{GeV}
\label{agecutoff}
\end{equation}
and
\begin{equation}
L_{\gamma}\sim 10^{36}K^3B_{d,12}^{-1/2}\tau_3^{-5/4}
~\mathrm{erg/s},
\label{agelumi}
\end{equation}
respectively,
 where $\tau_3$ is the spin down age in units of 1~kyrs and $L_{sd,
 34}$ is the spin down power in units of $10^{34}~\mathrm{erg/s}$.
Figures~\ref{age-cutoff} and~\ref{age-lumi} plot the cut-off energy and
the $\gamma$-ray luminosity as a function of the spin down age,
respectively.  The lines in
 Figure~\ref{age-cutoff} and Figure~\ref{age-lumi}
are results for $K=2$ and   the
 typical magnetic field of $<B_{d,12}>=3$
for the canonical pulsars (solid lines),  and for $K=15$  and
$<B_{d,12}>=3\times 10^{-4}$ for the  millisecond pulsars (dashed lines).
 We can see that the model predictions are consistent with the
observations for both populations. The present model predicts that the two
 populations will be separated in the plots of the emission properties
($E_c$, $L_{\gamma}$) as a function of the spin down age.

If we recast the cut-off energy $E_c$ given by equation~(\ref{cut-off}) and
the $\gamma$-ray luminosity $L_{\gamma}$ given by equation~(\ref{lumi}) 
as a function of the spin down power
$L_{sd}=(2\pi)^4B_d^2R_s^6/6c^3P^4$, we  obtain
\begin{equation}
E_{c}\sim 0.31K^{3/2}B_{d,12}^{1/4}L_{sd,34}^{1/4}~\mathrm{GeV}
\label{spcut}
\end{equation}
and
\begin{equation}
L_{\gamma}\sim 5\times 10^{32}K^3B_{d,12}^{3/4}L_{sd,34}^{5/8},
~\mathrm{erg/s}
\label{splumi}
\end{equation}
respectively.
One can see that the proportionality coefficients  $K^{3/2}B_{d,12}^{1/4}$
in equation~(\ref{spcut})
and  $K^3B_{d,12}^{3/4}$ in equation~(\ref{splumi})  have less dependency
on the pulsar populations.  For example, if we apply $K=2$ and
 the typical field  $<B_{d,12}>=3$ for the canonical pulsars and
$K=15$ and $<B_{d,12}>=3\times 10^{-4}$ for the millisecond pulsars, we
obtain $<K^{3/2}B_{d,12}^{1/4}>\sim3.7$
for the canonical pulsars and $\sim7.6$
for the millisecond pulsars,  and $<K^3B_{d,12}^{3/4}>\sim 18$ for the
canonical pulsars and $\sim 7.7$ for the millisecond pulsars.
We find that the proportionality coefficients between the two
populations are different only  about factor of two,
implying  it is  difficult to discriminate
between the two populations in the plot of the cut-off energy (or
$\gamma$-ray luminosity) versus  the spin down power.
 Figure~\ref{spin-cutoff} and~\ref{spin-lumi} plot the cut-off energy and
the $\gamma$-ray luminosity as a function of the spin down power,
respectively.  The solid lines are  model predictions with
$<K^{3/2}B^{1/4}_{d,12}>\sim5$ in Figure~\ref{spin-cutoff}
and $<K^3B_{d,12}^{3/4}>\sim 13$ in Figure~\ref{spin-lumi}.
Figures~\ref{spin-cutoff} and~\ref{spin-lumi} show
 the predicted relations that $E_{c}\propto L_{sd}^{1/4}$ and
$L_{\gamma}\propto L_{sd}^{5/8}$  are consistent with the observations
for both  canonical pulsars and  millisecond pulsars.

\subsection{Applicability to pulsars}
\label{limit}
In this section, we describe applicability of our model to the pulsars.  
The present gap closure model  invokes the condition that 
 the thickness $h_{\perp,m}$, above  which the magnetic pair-creation
 becomes to be important,  is less than the thickness, at which the gap is
 closed by the photon-photon pair-creation process.  To discuss the
 applicability to the pulsars, for example, one may compare 
the present model with a self-consistent model 
investigated by Zhang \& Cheng (1997), who
 discussed the gap thickness closed  by the photon-photon
 pair-creation process. They estimate the fractional gap thickness,
 which is defined by the ratio between  the typical gap
 thickness at $r=R_{lc}/2$  and the light radius,  as 
$f_{p}\equiv h_{\perp}(R_{lc/2})/R_{lc}\sim 5.5
P^{26/21}B_{d,12}^{-4/7}$. We recast our fractional gap thickness
defined in equation~(\ref{frac}) into one in
the sense  of the definition  by Zhang \& Chang (1997). 
 Because the trans-field thickness of the 
magnetic flux tube is approximately proportional to $h_{\perp}\propto
r^{3/2}$,  the fractional gap thickness is described by 
  $f_{m,1}\equiv h_{\perp,m}(R_{lc}/2)/R_{lc} 
\sim 2^{-3/2}f_m$, where $f_m$  is defined by equation~(\ref{frac}). 

 We compare the predicted thicknesses of $f_{p}$
 and $f_{m,1}$ for  36 canonical $\gamma$-ray pulsars in
 Table~1 and for 8 millisecond $\gamma$-ray pulsars in Table~2.
 We find that the pulsars with a larger spin down luminosity, such like
 PSR J0534+2200 (the Crab pulsar),  has $f_{p}<<f_{m,1}$. 
 As Table~1 and Table~2 show,  on the other hand, 
the mature pulsars, such like PSR J0633+1746 (the Geminga pulsar), 
 and the millisecond pulsars, who show a smaller spin down power, 
 indicate $f_{p}>>f_{m,1}$. 

The present gap closure process with the magnetic pair-creation process 
will be applicable for the pulsars whose have $f_{m,1}\la f_{p}$, that
is, the predicted gap thickness $f_{m,1}$ 
is comparable with or slightly less than $f_{p}$ of the
photon-photon pair-creation process.  
The condition that  $f_{p}<<f_{m,1}$ implies that   the 
 efficient photon-photon pair-creation will definitely close the gap 
before the gap
 reaches  the thickness $f_{m,1}$ . 
For the case $f_{p}>>f_{m,1}$, on the other hand, 
 the outer gap must be  thicker  than $f_m$, but 
the gap may be closed by magnetic pair-creation process with
 the thickness at which the photon-photon pair-creation is initiated in
 the gap.  Equating the fractional gap thickness $f_p$ and $f_{m,1}$,
 it may be suggested that the switching of the gap closure mechanism
 takes palace at 
\begin{equation}
L_{sd}\sim 2.5\times
 10^{36}(K/2)^{-168/31}B_{d,12}^{-34/31}~\mathrm{erg/s}, 
\end{equation}
indicating the photon-photon pair-creation closes the gap 
for the pulsars with $L_{sd}\ga 10^{36}~\mathrm{erg/s}$, 
while the magnetic pair-creation may close the 
gap for $L_{sd}\la 10^{36}~\mathrm{erg/s}$.

\subsection{Validity of assumption}
\label{valid}
In the present model, we assume that  the energy loss due to the curvature
radiation below  the null charge surface  is not compensated
by the acceleration of the electric field in the gap.  For simplicity,
we assume that
the electric field decreases  quadratically along the field lines below
 the null charge surface toward  the inner boundary;
\begin{equation}
E_{||}(r<R_{n})=\frac{(r/R_{in})^2-1}{(R_n/R_{in})^2-1} E_{||,0},
\end{equation}
where $R_{in}$ and $R_n$ are the radial distance to the inner boundary and
the null charge point, respectively, and $E_{||,0}$ is the electric
field at the null charge point, which is characterized by
equation~(\ref{accfield}).  If the inner boundary is
located $\sim 1$ stellar radius from the stellar surface,  the acceleration
field is reduced  to $E_{||}(r\sim R_{in})\sim (R_{in}/R_n)^2E_{||,0}\sim
10^{-2}E_{||,0}$ near the
inner boundary. Applying the typical electric field described by
equation~(\ref{accfield}) at the null charge point,
 a  particle gains energy by  the electric
field  with a rate of
\begin{equation}
eE_{||}(r\sim R_{in})c\sim 40K^2 B_{d,12}P^{-2}_{-1}~\mathrm{erg/s}.
\end{equation}
On the other hand, the particles having
the Lorentz factor expressed by equation~(\ref{lorent})  loose their
energy via the curvature radiation with a rate of
\begin{equation}
P_c=\frac{2}{3}\gamma^4\frac{e^2c}{s^2}\sim 5\times 10^3s_7^{-2/3}
~\mathrm{erg/s},
\end{equation}
indicating the energy  loss by the curvature radiation
 dominates the energy gain by the acceleration of the electric
 field. In fact, (1)  the electrodynamic model expects
 the electric field will decreases more rapidly below the null charge
surface  than the quadratic expression (Hirotani 2006a) or  (2) the
 inner boundary will located far from the stellar surface for the small
 current case.  Therefore we can safely assume that
the energy loss due to the curvature
radiation below  the null charge surface  is not compensated
by the acceleration of the electric field.

\subsection{Comparison with the previous works}
We briefly summarize the difference in the predicted
$\gamma$-ray luminosity  between
the present model and the model investigated by Zhang \& Cheng~(1997).
Zhang\& Cheng~(1997) predicted the fractional gap thickness
$f_{p}\sim  5.6P^{26/21}B_{d,12}^{-4/7}$ of the outer gap closed  by
the photon-photon pair-creation process in the outer magnetosphere. With
this model, the $\gamma$-ray luminosity depends on the spin down
power as $L_{\gamma}\propto B_d^{1/7}L_{sd}^{1/14}$. Therefore,
the $\gamma$-ray luminosity is less dependent on the spin down
power.  Later, they developed their gap model by taking into
 account the effects of the
inclination angle between the rotation axis and
the magnetic axis (Zhang et al. 2004).
They argued that the fractional gap thickness and the resultant  $\gamma$-ray
luminosity depend on the inclination angle. They assumed that
when the fractional gap thickness  goes to unity, then the $\gamma$-ray
luminosity approaches to  the spin down power. Therefore, the
$\gamma$-ray luminosity bounded between $L_{sd}^{1/14}$ and $L_{sd}$
depending on the inclination angle.
They carried out a Monte Carlo simulating  to calculate the $\gamma$-ray
luminosity by assuming that the inclination angle is randomly
distributed. As a result, they obtained the relation  that
$L_{\gamma}\propto L_{sd}^{\beta}$ with an index of
$\beta=0.38\sim 0.46$. In the present model, on the other hand,
 the fractional gap thickness $f_m\propto P^{1/2}$ has a  less dependency
 on the rotational period compared with the model of Zhang \& Cheng
 (1997), in which $f_p\propto P^{26/21}$.
As a result, without introducing  the effects of the inclination angle,
the present model  predicts  the relation of
$L_{\gamma}\propto  L_{sd}^{\beta}$ with more  steep index of $\beta=5/8$.

\section{Conclusion}
\label{conclusion}
\begin{figure}
\epsscale{.50}
\plotone{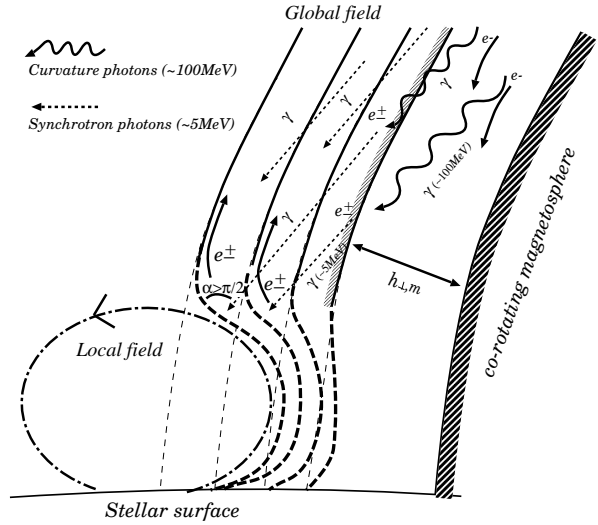}
\caption{Schematic view of the favorable magnetic structure near the
 stellar surface.}
\label{Pulsar}
\end{figure}

\begin{figure}
\epsscale{.50}
\plotone{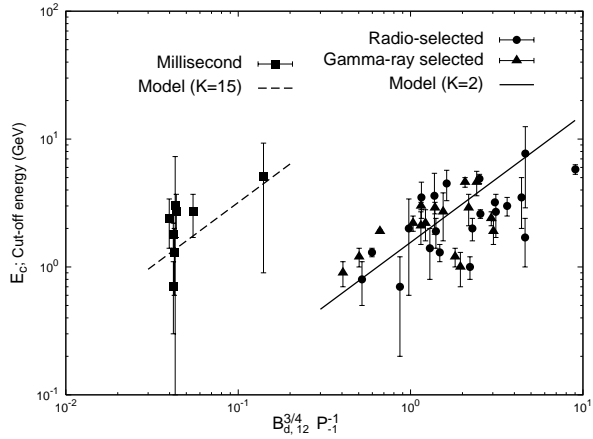}
\caption{Plot of the cut-off energy as a function of
 $B_{d,12}^{3/4}P_{-1}^{-1}$. The circle and triangle symbols are observations
for the radio-selected and the $\gamma$-ray selected $\gamma$-ray
 pulsars, respectively. In addition,  the square  symbols  represent the
 millisecond pulsar. The solid and dashed lines show the prediction of
 the model, $E_{c}=0.55K^{3/2}B_{d,12}^{3/4}P^{-1}_{-1}~\mathrm{GeV}$ of
 equation (\ref{cut-off}). The model results are for
 $K=2$ for the canonical pulsar (solid line) and $K=15$
for the millisecond
 pulsars (dashed line). The observations are taken from Abdo et al. (2010).
}
\label{cutoff}
\end{figure}

\begin{figure}
\epsscale{.50}
\plotone{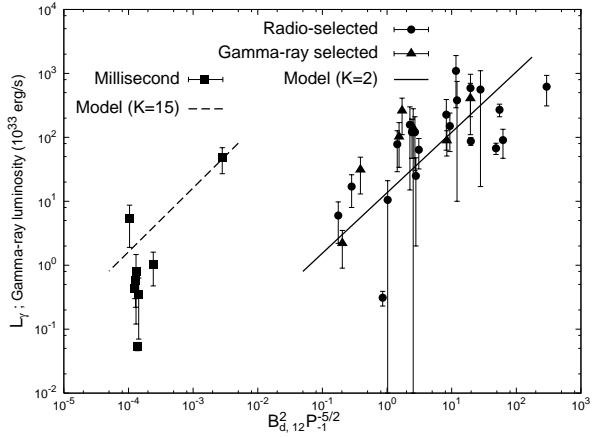}
\caption{
Plot of the $\gamma$-ray luminosity  as a function of
 $B_{d,12}^{2}P_{-1}^{-5/2}$. The symbols correspond to same case
as Figure~\ref{cutoff}. The lines represent the model prediction,
$L_{\gamma}\sim 2\times
 10^{33}K^3B_{d,12}^2P^{-5/2}_{-1}~\mathrm{erg/s}$ of equation~(\ref{lumi}),
with $K=2$ for the canonical pulsar (solid line)
and $K=15$ for the millisecond pulsars (dashed line).
}
\label{lumin}
\end{figure}

\begin{figure}
\epsscale{.50}
\plotone{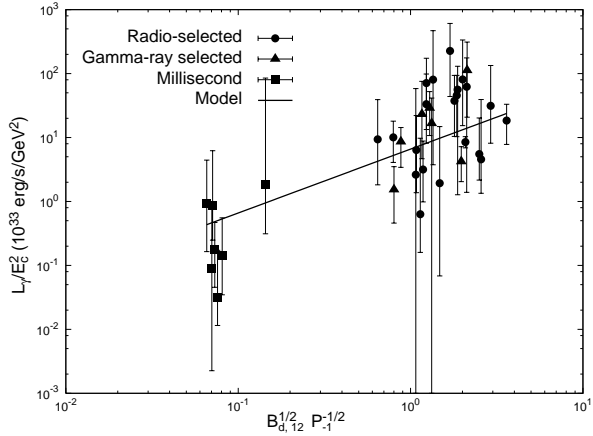}
\caption{
Plot of $L_{\gamma}/E_c^2$ as a function of
 $B_{d,12}^{1/2}P_{-1}^{-1/2}$. The symbols correspond to same case
as Figure~\ref{cutoff}. The line corresponds to the model prediction,
 $L_{\gamma}/E_{c}^2\sim 6.6\times 10^{34}B_{d,12}^{1/2}P_{-1}^{-1}
~\mathrm{erg/s~GeV^2}$ of equation~(\ref{basiceq}).
}
\label{basic}
\end{figure}

\begin{figure}
\epsscale{.50}
\plotone{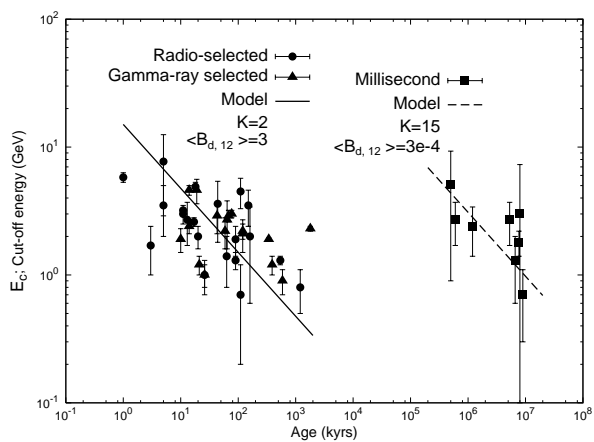}
\caption{Plot of the cut-off energy as a fucntion of the spin down
 age. The symbols correspond to same case
as Figure~\ref{cutoff}.  The lines correspond to the model
prediction,
$E_{c}\sim 7K^{3/2}B_{d,12}^{-1/4}
\tau_3^{-1/2}~\mathrm{GeV}$ of equation (\ref{agecutoff}),  with
$K=2$ and the typical magnetic field $<B_{d,12}>=3$
for the canonical pulsars (solid line), and $K=15$ and
 $<B_{d,12}>=3\times 10^{-4}$
 for the millisecond pulsars (dashed line). }
\label{age-cutoff}
\end{figure}

\begin{figure}
\epsscale{.50}
\plotone{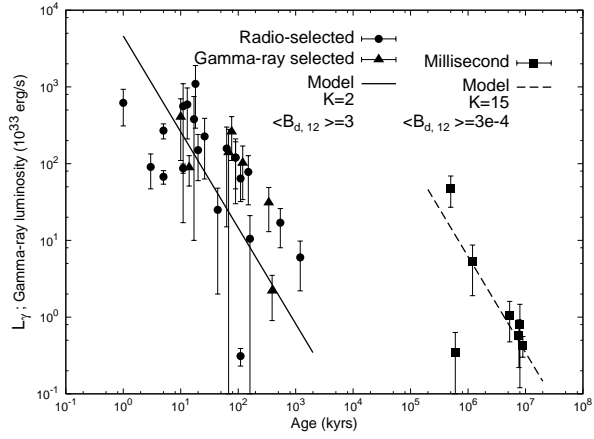}
\caption{Plot of the $\gamma$-ray luminosity as a function of the spin down
 age. The symbols  correspond to same case
as Figure~\ref{cutoff}. The lines correspond to the model prediction,
 $L_{\gamma}\sim 10^{36}K^3B_{d,12}^{-1/2}\tau_3^{-5/4}~~\mathrm{erg/s}$
 of equation (\ref{agelumi}),  with
$K=2$ and  $<B_{d,12}>=3$
for the canonical pulsars (solid line), and $K=15$ and
 $<B_{d,12}>=3\times 10^{-4}$
 for the millisecond pulsars (dashed line). }
\label{age-lumi}
\end{figure}

\begin{figure}
\epsscale{.50}
\plotone{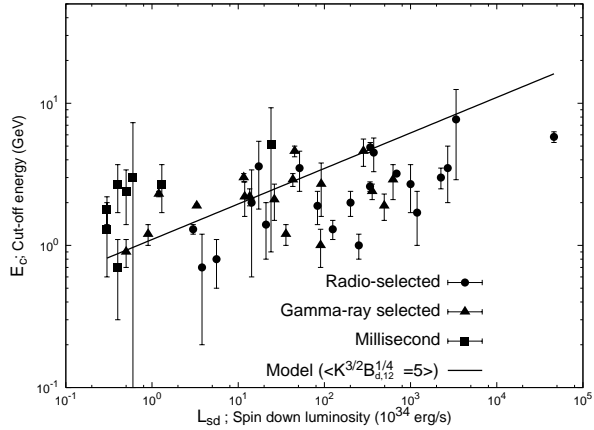}
\caption{Plot of the cut-off energy as a fucntion of the spin down
 power. The symbols correspond to same case
as Figure~\ref{cutoff}.  The line is the model prediction,
$E_{c}\sim
 0.22K^{3/2}B_{d,12}^{1/4}L_{sd,34}^{1/4}~\mathrm{GeV}$ of equation 
(\ref{spcut}), with $<K^{3/2}B^{1/4}_{d,12}>=5$.}
\label{spin-cutoff}
\end{figure}

\begin{figure}
\epsscale{.50}
\plotone{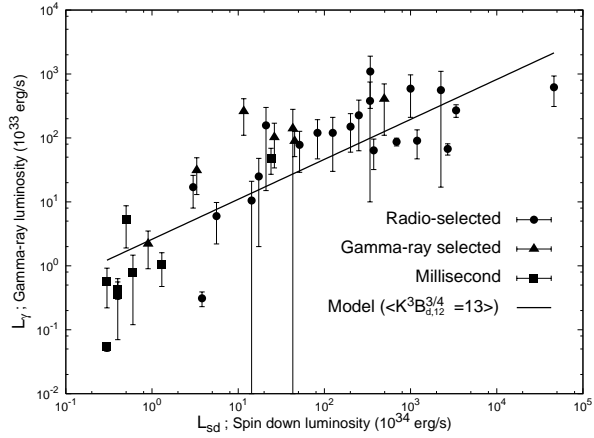}
\caption{Plot of the $\gamma$-ray luminosity as a function of the spin down
 power. The symbols correspond to same case
as Figure~\ref{cutoff}.  The line is the model prediction,
$L_{\gamma}\sim 2\times
 10^{32}K^3B_{d,12}^{3/4}L_{sd,34}^{5/8}~\mathrm{erg/s}$ of equation 
(\ref{splumi}), with
$<K^{3}B^{3/4}_{d,12}>=13$.}
\label{spin-lumi}
\end{figure}

In this paper, we have studied  the outer gap accelerator
model closed  by magnetic pair-creation process.
We argued  that  below null charge surface, the curvature loss is not
compensated by the acceleration due to the electric field in the gap.
In such a case, the incoming particles, which were produced in the outer
magnetosphere,   will emit curvature photons with
about $E_{min}\sim m_ec^2/\alpha_{f}\sim 100$~MeV.
 The 100~MeV curvature photos propagating toward the stellar surface will be
 converted into pairs by the pair-creation process with the strong local
 magnetic field near the stellar surface, where the multiple magnetic
 field affects to the global field lines.  For the canonical pulsar, 
the synchrotron radiation of the created pairs produce  $\sim$5~MeV photon, 
which will be furthermore converted into pairs.  As a result, 
multiplicity of an incoming
 particle could acquire $M_{e^{\pm}}\sim 10^4-10^5$.  With the 
local field lines bending away
from the last-open field line (such as illustrated in
Figure~\ref{Pulsar}),  the created pairs via the magnetic pair-creation 
process can have the outgoing  momentum and migrate into the outer 
magnetosphere. If $\sim 10$ pairs out of  $M_{e^{\pm}}\sim 10^4-10^5$
migrate  outward to 
 the outer magnetosphere, those pairs 
could close  the outer gap accelerator. 
  According to this scenario,
the main results of this paper are as follows.
The fractional thickness of the outer gap
becomes $f_m\sim 0.25K P^{1/2}_{-1}$, which has a less dependency on
 the rotational period compared with the outer gap model proposed by
 Zhang \& Cheng (1997).  With the present model, the spectral properties
of the $\gamma$-ray emissions depend on the local parameter
$K\sim \chi_{-1}^2B^{-2}_{m,12}s_7(R_s/R_i)^{3/2}$, which
is determined by the local magnetic structure near the star. 
We expect that the local parameter $K$ takes a vale of $K\sim
2$  for  the canonical pulsars and $K\sim 15$ for the millisecond
pulsars (Figures~\ref{cutoff} and~\ref{lumin}). The present model
predicts that the canonical pulsars and the millisecond pulsars are
connected in the plots of $L_{\gamma}/E_{c}$ versus $B_{s}^{1/2}P^{-1/2}$,
in which the effect of the fractional gap thickness is carried away
(Figure~\ref{basic}).  The present model predicts that the cut-off
energy ($E_c$) and the $\gamma$-ray luminosity ($L_{\gamma}$)
depend on the spin down age or
the spin down power  as $E_{c}\propto \tau^{-1/2}$ and
$L_{\gamma}\propto \tau^{-5/4}$ or  $E_{c}\propto L_{sd}^{1/4}$ and
$L_{\gamma}\propto L_{sd}^{5/8}$ (Figures~\ref{age-cutoff}-\ref{spin-lumi}).

In addition to the cut-off energy and $\gamma$-ray luminosity, which
have been discussed in this paper,  the
$Fermi$ $\gamma$-ray telescope provides  the photon index of
$\gamma$-ray spectrum  and the pulse profiles for each pulsar (Abdo et
al. 2010). It must be  important to discuss the photon index and the
pulse profile with the acceleration model, because they will contain
information of the electric
structure (e.g. the distribution of the electric field)
in the acceleration region and the three-dimensional geometry of the
emission region (e.g. Romani \& Yadigaroglu 1995; Cheng, Ruderman \&
Zhang 2000; Spitkovsky 2006). However,
 a more  detail  model, which has to consider the electrodynamics  in the gap
and three-dimensional structure, is required to study  the shape of the
$\gamma$-ray spectra and the pulse profiles.  Studying the emission
properties with the electrodynamics in the present outer gap closure
model will be done in the subsequent papers.

\acknowledgments
We wish to express our thanks to the referee for insightful comments on
the manuscript. We thank the useful discussions with  H.-K. Chang,
K.~Hirotani, C.Y.~Hui, B.~Rudak,
M.Ruderman and  S.Shibata. We also thank to Theoretical Institute
for Advanced Research in Astrophysics (TIARA) operated under Academia
Sinica Institute of Astronomy and Astrophysics, Taiwan,
which  enables author (J.T.) to use PC cluster at TIARA.
This work is supported by a GRF grant of Hong Kong SAR
Government under HKU700908P.





\clearpage


\begin{table}
\begin{tabular}{cccccc}
\hline\hline
Name & $P(s)$ & $B_d (10^{12}~\mathrm{G})$ & $L_{sd} (10^{34}~\mathrm{erg/s})$ & $f_p$ & $f_{m,1} ; K=2$ \\
\hline
 J0534$+$2200  & 0.033 & 3.8 & 46100 & 0.038 & 0.10 \\
J1833$-$1034  & 0.062 & 3.5 & 3370 & 0.085 & 0.14 \\
J0205$+$6449  & 0.066 & 3.6 & 2700 & 0.091 & 0.14 \\
J2229$+$6114  & 0.052 & 2.0 & 2250 & 0.093 & 0.13 \\
J1124$-$5916  & 0.135 & 10.0 & 1190 & 0.12 & 0.21 \\
J1420$-$6048  & 0.068 & 2.4 & 1000 & 0.12 & 0.15 \\
J0835$-$4510  & 0.089 & 3.4 & 688 & 0.14 & 0.17 \\
J1813$-$1246  & 0.048 & 0.9 & 626 & 0.13 & 0.12 \\
J1418$-$6058  &  0.11  & 4.4 & 495 & 0.16 & 0.19 \\
 J1952$+$3252  & 0.040 & 0.5 & 374 & 0.15 & 0.11 \\
J1826$-$1256  & 0.11  & 3.7 & 358 & 0.17 & 0.19 \\
J1709$-$4429  & 0.102 & 3.0 & 341 & 0.17 & 0.18 \\
 J2021$+$3651  & 0.104 & 3.2 & 338 & 0.17 & 0.18 \\
 J1907$+$06    & 0.107 & 3.1 & 284 & 0.18 & 0.18 \\
J1747$-$2958  & 0.990 & 2.5 & 251 & 0.19 & 0.18 \\
J1048$-$5832  & 0.124 & 3.5 & 201 & 0.20 & 0.20 \\
 J1718$-$3825  & 0.075 & 1.0 & 125 & 0.22 & 0.15 \\
J1459$-$60    & 0.103 & 1.6 & 91.9 & 0.25 & 0.18 \\
J2238$+$59    & 0.163 & 4.1 & 90.3 & 0.26 & 0.23 \\
J1028$-$5819  & 0.091 & 1.2 & 83.2 & 0.25 & 0.17 \\
J1509$-$5850  & 0.089 & 0.9 & 51.5 & 0.29 & 0.17 \\
J0007$+$7303  & 0.316 & 10.6 & 45.2 & 0.34 & 0.31 \\
J1809$-$2332  & 0.147 & 2.2 & 43 & 0.32 & 0.21 \\
 J1958$+$2846  & 0.29  & 8.0 & 35.8 & 0.36 & 0.30 \\
\hline
\end{tabular}
\end{table}

\begin{table}
\begin{tabular}{cccccc}
\hline\hline
Name & $P(s)$ & $B_d (10^{12}~\mathrm{G})$ & $L_{sd} (10^{34}~\mathrm{erg/s})$ & $f_p$ & $f_{m,1} ; K=2$ \\
\hline
  J2032$+$4127  & 0.143 & 1.7 & 26.3 & 0.37 & 0.21 \\
J0248$+$6021  & 0.217 & 3.4 & 21 & 0.41 & 0.26 \\
J0631$+$1036  & 0.288 & 5.5 & 17.3 & 0.45 & 0.30 \\
J0742$-$2822  & 0.167 & 1.7 & 14.3 & 0.45 & 0.23 \\
J1732$-$31    & 0.197 & 2.2 & 13.6 & 0.46 & 0.25 \\
J0633$+$0632  & 0.297 & 4.8 & 11.9 & 0.50 & 0.30 \\
J2021$+$4026  & 0.265 & 3.8 & 11.6 & 0.50 & 0.29 \\
J2043$+$2740  & 0.096 & 0.3 & 5.6 & 0.55 & 0.17 \\
J0659$+$1414 & 0.385 & 4.3 & 3.8 & 0.73 & 0.35 \\
J0633$+$1746  & 0.237 & 1.6 & 3.3 & 0.71 & 0.27 \\
J1057$-$5226  & 0.197 & 1.1 & 3 & 0.70 & 0.25 \\
J1836$+$5925  & 0.173 & 0.5 & 1.2 & 0.92 & 0.23 \\
J1741$-$2054  & 0.414 & 2.3 & 0.9 & 1.14 & 0.36 \\
J0357$+$32    & 0.444 & 1.9 & 0.5 & 1.40 & 0.37 \\
\hline
\end{tabular}
\caption{Pulsar parameters, which were  taken from Abdo et al. (2010), 
 and the fractional gap thickness 
predicted by Zhang \& Cheng (1997), $f_{p}$, and by the
 present model, $f_{m,1}$, for 38 canonical $\gamma$-ray pulsars. Here, 
the fractional gap thickness is defined by
 the ratio between the typical gap thickness at $r=R_{lc}/2$ and  the
 light radius. }
\end{table}

\begin{table}
\begin{tabular}{cccccc}
\hline\hline
Name & $P(ms)$ & $B_d (10^{8}~\mathrm{G})$ & $L_{sd} (10^{34}~\mathrm{erg/s})$ & $f_p$ & $f_{m,1} ; K=15$ \\
\hline
J0218$+$4232  & 2.3 & 4.1 & 24 & 0.26 & 0.20 \\
J0613$-$0200  & 3.1 & 1.8 & 1.3 & 0.60 & 0.23 \\
J0751$+$1807  & 3.5 & 1.5 & 0.6 & 0.76 & 0.25 \\
J1614$-$2230  & 3.2 & 1.2 & 0.5 & 0.78 & 0.24 \\
J1744$-$1134  & 4.1 & 1.8 & 0.4 & 0.84 & 0.27 \\
J2124$-$3358  & 4.9 & 2.4 & 0.4 & 0.89 & 0.29 \\
J0030$+$0451  & 4.9 & 2.2 & 0.3 & 0.92 & 0.29 \\
 J0437$-$4715  & 5.8 & 2.9 & 0.3 & 0.98 & 0.32 \\
\hline
\end{tabular}
\caption{Pulsar parameters, which were taken from Abdo et al. (2010), 
 and the fractional gap thickness 
predicted by Zhang \& Cheng (1997), $f_{p}$, and by the
 present model, $f_{m,1}$, for 8 millisecond  $\gamma$-ray pulsars. Here, 
the fractional gap thickness is defined by
 the ratio between the typical gap thickness at $r=R_{lc}/2$ and  the
 light radius. }
\end{table}

\newpage

\newpage


\begin{thebibliography}{}

\bibitem[\protect\citeauthoryear{Abdo}{2010}]{ab010}
Abdo A.A. et al., 2010, ApJS,  187, 460
\bibitem[\protect\citeauthoryear{Abdo}{2009}]{ab09a}
Abdo A.A. et al., 2009b, Sci., 325, 848
\bibitem[\protect\citeauthoryear{Abdo}{2009}]{ab09b}
Abdo A.A. et al., 2009c, Sci., 325, 840
\bibitem[\protect\citeauthoryear{Abdo}{2009}]{ab09c}
Abdo A.A. et al., 2009d, ApJ, 696, 1084
\bibitem[\protect\citeauthoryear{Aliu}{2009}]{al09}
Aliu E. et al., 2008, Sci, 322, 1221
\bibitem[\protect\citeauthoryear{Arons}{1993}]{ar93}
Arons J., 1993, ApJ, 408, 160
\bibitem[\protect\citeauthoryear{Arons}{1983}]{ar83}
Arons J., 1983, ApJ, 266, 215
\bibitem[\protect\citeauthoryear{Blandford}{1983}]{bl83}
Blandford R.D.,  Applegate J.H. \&  Hernquist, L., 1983, MNRAS, 204, 1025
\bibitem[\protect\citeauthoryear{Cheng}{2000}]{ch00}
Cheng K.S., Ruderman M. \& Zhang L. 2000, ApJ, 537, 964
\bibitem[\protect\citeauthoryear{Cheng}{1986a}]{ch86a}
 Cheng K.S., Ho C. \&  Ruderman M. 1986a, ApJ, 300, 500
\bibitem[\protect\citeauthoryear{Cheng}{1986b}]{ch86b}
 Cheng K.S., Ho C. \& Ruderman M. 1986b, ApJ, 300, 522
\bibitem[\protect\citeauthoryear{Goldreich}{1969}]{go69}
Goldreich P. \&  Julian W.H. 1969, ApJ, 157, 869
\bibitem[\protect\citeauthoryear{Daugherty}{1996}]{da96}
Daugherty J.K. \&  Harding, A.K., 1996, ApJ, 458, 278
\bibitem[\protect\citeauthoryear{Daugherty}{1982}]{da82}
Daugherty J.K. \&  Harding, A.K., 1982, ApJ, 252, 337
\bibitem[\protect\citeauthoryear{Dyks}{2003}]{dy03}
Dyks J. \& Rudak B.,  2003, ApJ, 598, 1201
\bibitem[\protect\citeauthoryear{Dyks}{2004}]{dy04}
Dyks J., Rudak B. \& Harding A.K.,  2004, ApJ, 607, 939
\bibitem[\protect\citeauthoryear{Erber}{1966}]{er66}
Erber T.,  1966, RvMP, 38, 626
\bibitem[\protect\citeauthoryear{Harding}{2008}]{ha08}	
Harding A.K., Stern J.V., Dyks J. \&  Frackowiak M., 2008, ApJ, 680, 1378
\bibitem[\protect\citeauthoryear{Harding}{2005}]{ha05}
Harding A.K., Usov V.V., Muslimov A.G., 2005, ApJ, 622, 531
\bibitem[\protect\citeauthoryear{Hirotani}{2003}]{hi03}
Hirotani K., Harding A.K. \& Shibata S., 2003, ApJ, 591, 334
\bibitem[\protect\citeauthoryear{Hirotani}{2008}]{hi08}
Hirotani K., 2008, ApJL, 688, 25
\bibitem[\protect\citeauthoryear{Hirotani}{2006a}]{hi06a}
Hirotani K., 2006a, ApJ, 652, 1475
\bibitem[\protect\citeauthoryear{Hirotani}{2006b}]{hi06b}
Hirotani K., 2006b, Mod. Phys. Lett. A, 21, 1319
\bibitem[\protect\citeauthoryear{Hirotani}{2001}]{hi01}
Hirotani K. \& Shibata S., 2001, ApJ, 558, 216H	
\bibitem[\protect\citeauthoryear{Pellizzoni}{2009}]{pe09}
Pellizzoni, A. et al., 2009, ApJ, 695, 115
\bibitem[\protect\citeauthoryear{Romani}{1995}]{ro95}
  Romani R.W. \& Yadigaroglu I.-A. 1995, ApJ, 438, 314
\bibitem[\protect\citeauthoryear{Romani}{1990}]{ro90}
Romani R.W, 1990, Nature, 347, 741
\bibitem[\protect\citeauthoryear{Ruderman}{1991}]{ru91}
Ruderman M., 1991, ApJ, 366, 261	
\bibitem[\protect\citeauthoryear{Ruderman}{1975}]{ru75}
Ruderman M.A. \& Sutherland P.G., 1975, ApJ, 196, 51
\bibitem[\protect\citeauthoryear{Spitkovsky}{2006}]{spi06}
Spitkovsky A., 2006, ApJL, 648L, 51
\bibitem[\protect\citeauthoryear{Tang}{2008}]{tan08}
Tang P.S. Anisia, Takata J., Jia, J.J., Cheng K.S., 2008, ApJ, 676, 562
\bibitem[\protect\citeauthoryear{Takata}{2009}]{ta09}
Takata J. \& Chang H.-K., 2009 MNRAS, 392, 400
\bibitem[\protect\citeauthoryear{Takata}{2007}]{ta07a}
Takata J. \& Chang H.-K., 2007 ApJ, 670, 677
\bibitem[\protect\citeauthoryear{Takata}{2007}]{ta07b}
Takata J., Chang H.-K. \& Cheng K.S., 2007 ApJ, 656, 1044
\bibitem[\protect\citeauthoryear{Takata}{2004}]{ta04}
Takata J., Shibata S. \&  Hirotani K. 2004, MNRAS, 354, 1120
\bibitem[\protect\citeauthoryear{Thompson}{2003}]{th02}
Thompson D.J., 2004, in Cheng K.S., Romero G.E., eds, Cosmic Gamma Ray
Sources. Dordrecht, Kluwer, p. 149
\bibitem[\protect\citeauthoryear{Venter}{2009}]{ve09}
Venter C.,  Harding A.K. \& Guillemot L., 2009, ApJ, 707, 800
\bibitem[\protect\citeauthoryear{Zhang}{2004}]{za04}
Zhang L., Cheng K.S., Jiang Z.J. \&  Leung P., 2004, ApJ, 604, 317
\bibitem[\protect\citeauthoryear{Zhang}{2003}]{za03}
Zhang L. \&  Cheng K.S.,  2003, A\& A 398, 639	
\bibitem[\protect\citeauthoryear{Zhang}{1997}]{za97}
Zhang L. \&  Cheng K.S.,  1997, ApJ, 487, 370
\end{thebibliography}
\end{document}